\documentclass{ws-p8-50x6-00}
\newcommand{\gev}{\mbox{\,Ge\kern-0.2exV}}
\newcommand{\mev}{\mbox{\,Me\kern-0.2exV}}

\def\ifmath#1{\relax\ifmmode #1\else $#1$\fi}%
\def\rA{\ifmath{{\mathrm{A}}}}
\def\rC{\ifmath{{\mathrm{C}}}}
\def\rF{\ifmath{{\mathrm{F}}}}
\def\rb{\ifmath{{\mathrm{b}}}}
\def\rd{\ifmath{{\mathrm{d}}}}
\def\rg{\ifmath{{\mathrm{g}}}}
\def\rp{\ifmath{{\mathrm{p}}}}
\def\rt{\ifmath{{\mathrm{t}}}}

\def\rq{\ifmath{{\mathrm{q}}}}
\def\Le{\ifmath{{\mathrm{Le}}}}
\def\Lu{\ifmath{{\mathrm{Lu}}}}

\begin{document}

\title{Test of QCD Predictions for\\ Multiparticle Production at LEP}
\author{O. Passon}
\address{Fachbereich Physik, Bergische Universit{\"a}t-GH Wuppertal,
 Gau\ss{}stra\ss{}e 20, \\ D-42097 Wuppertal,
Germany\\E-mail: Oliver.Passon@CERN.CH}


\maketitle

\abstracts{I discuss various QCD tests for multiparticle production, such as 
multiplicities, dead-cone effect and inclusive spectra. The common feature of
all these predictions is the crucial importance of coherence effect to be
taken into account.}

\section{Charged Particle Multiplicities}
\begin{figure}[ht]
\begin{center}
 \vspace{-0.5cm}
\epsfxsize=13pc % will enlarge or reduce the postscript figures based on t1he 
\epsfbox{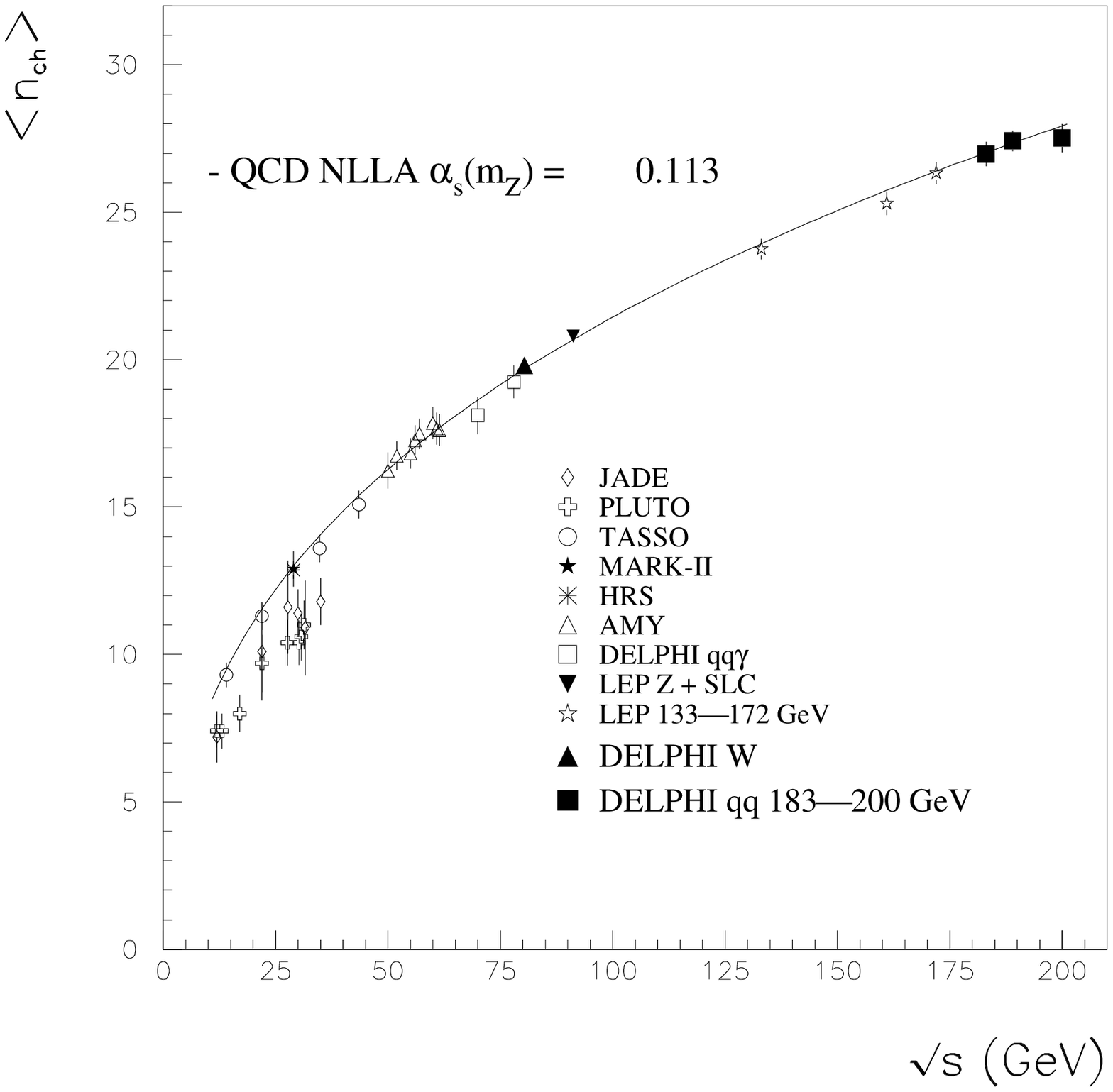} % postscript image file name
\epsfxsize=13pc
\epsfbox{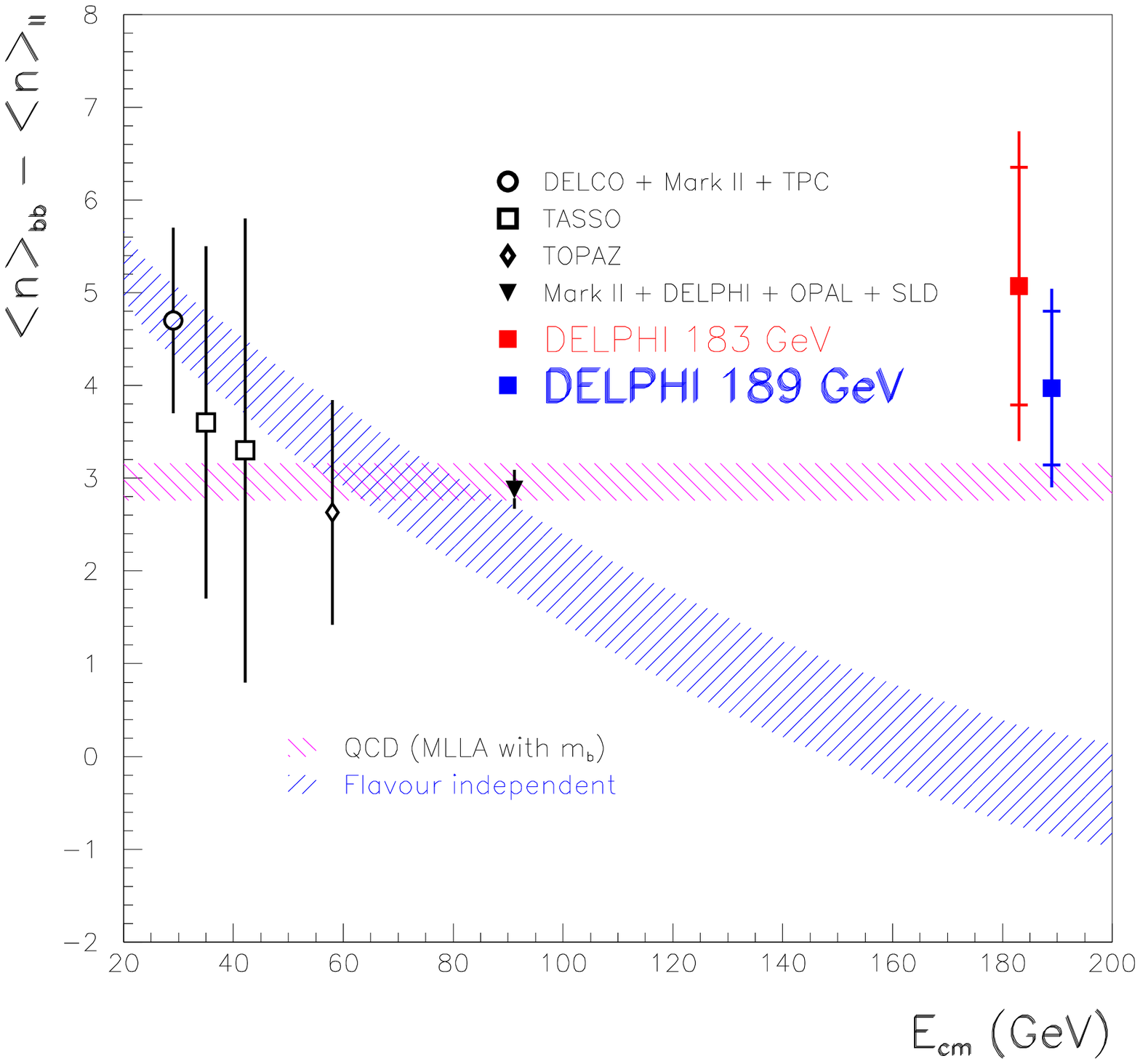} % postscript image file name
\caption{ a) Charged multiplicities at various $E_{\mathrm CM}$ up to
 200 \gev\ b) Energy dependence of the differnce between mean multiplicity for
 $\rb\bar{\rb}$ and light quark events.\label{multi}}
\end{center}
\end{figure}

Among the most general features of e$^+$e$^-$ annihilation one can look at is
the charged particle multiplicity. Its mean value is predicted as a function 
of two free parameters: $\alpha_s$ and the constant $a$ in the following 
formula:
$$
\langle n_{ch} \rangle (Q)=a \alpha_s(Q)^b \exp{c/\sqrt{\alpha_s}}[1+{\cal O}
(\sqrt{\alpha_s})] 
$$
The constants $b$ and $c$ entering in this expression can be calculated, and 
especially 
in the result for $c$ enters crucially the assumption of coherence: neglecting
the effect of angular ordering would increase it by a factor of $\sqrt{2}$.
The data up to the highest energies keep to be in fair agreement with 
this prediction \cite{delphi1} (see Fig.\ref{multi}). 
An important point in this measurement is that 
the LEP data have been corrected for the different b quark fractions.
%Of course the charged multiplicity is not just a number, but a distribution, 
%thus 
%%is is not characterised my a mean value only, but also a width (among other 
%higher moments).

The ratio of mean and dispersion is predicted
to have a very mild energy dependence, or even to be energy independent 
according  to the KNO scaling property. Also these predictions are met by 
the data within errors.\cite{delphi1}
%%%%%%%%%%%%%%%%%%%%%%%%%%%%%%%%%%%%%%%%%%%%%%%%%%%%%%%%%%%%%%%%%%%%%%%
\section{Dead Cone Effect}
It was mentioned above that the mean multiplicities were corrected for 
the different b quark fraction, since {\em this} QCD calculation treats 
quarks as massless particles. A prominent example for a mass effect as 
predicted by QCD is the so-called dead-cone-effect, which originates 
from the fact that radiation from a
 massive object is suppressed in a cone with half opening angle $m/E$.    
Evidently, this has the strongest effect on the $\approx$ 5 \gev\ heavy b 
quark, but happens to be still hard to detect directly. As a 
consequence of the
suppressed radiation, the difference between the mean multiplicity of b and
light quark events is predicted to be energy independent.\cite{schumm}
Fig.\ref{multi} b) shows this multiplicity difference as
 a function of energy.\cite{delphi2}
Indeed the LEP2 data confirm the coherent scenario, while in the naive
 approach one expects a decreasing influence of any mass effect in the limit 
of higher energies. 
%%%%%%%%%%%%%%%%%%%%%%%%%%%%%%%%%%%%%%%%%%%%%%%%%%%%%%%%%%%%%%%%%%%%
\section{Inclusive Spectra}
Let us now turn back to a more general feature, the inclusive spectra of all 
charged particles.
The textbook example for the analytic perturbative approach is the 
$\xi_p$ distribution, and the energy dependence of its maximum, $\xi^*$. 
In the
limited spectrum approximation the distribution is predicted as a function 
of two free
paramters: $\Lambda_{\mathrm eff}$ and the normalization $K$, which relates
 parton and hadron level. For the maximum $\xi^*$ the $K$ dependence drops 
out evidently.
\begin{figure}[ht]
 \begin{center}
  \vspace{-0.1cm}
  \unitlength1cm
  \mbox{\epsfig{file=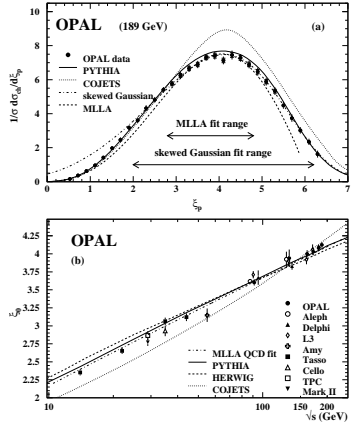,width=5cm}}
 \end{center}
  \vspace{-0.5cm}
\caption{\label{xi} $\xi_p$ distribution and energy dependence of its maximum,
 $\xi^*$.}
\end{figure}  
Fig.\ref{xi} shows the distribution as measured by OPAL \cite{opal} at
 189 \gev. The 
dashed line is a fit of the MLLA calculation to the data where the value of 
$\Lambda_{\mathrm eff}$ was fixed at 250 \mev\ and only the normalization $K$ 
was fitted. As for the maximum $\xi^*$, only the incoherent COJET model 
is unable to  describe the situation properly. It should be noted, however,
that the assumed energy independence of the normalization $K$ is only observed
 on a 10\% level.\cite{opal} 
\begin{figure}[hb]
\begin{center}
 \vspace{-0.5cm}
\epsfxsize=12pc % will enlarge or reduce the postscript figures based on t1he 
\epsfbox{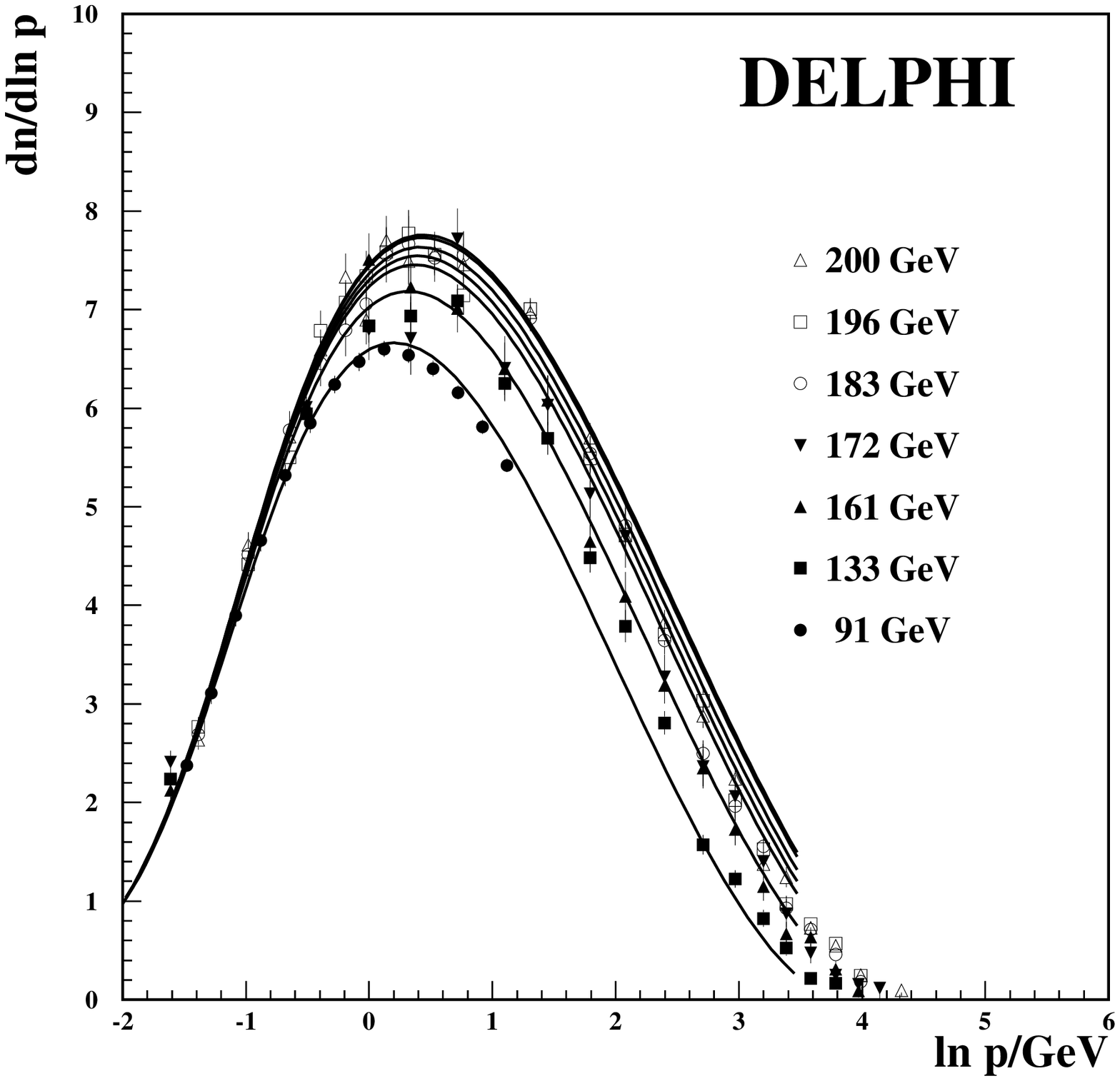} % postscript image file name
\epsfxsize=11pc
\epsfbox{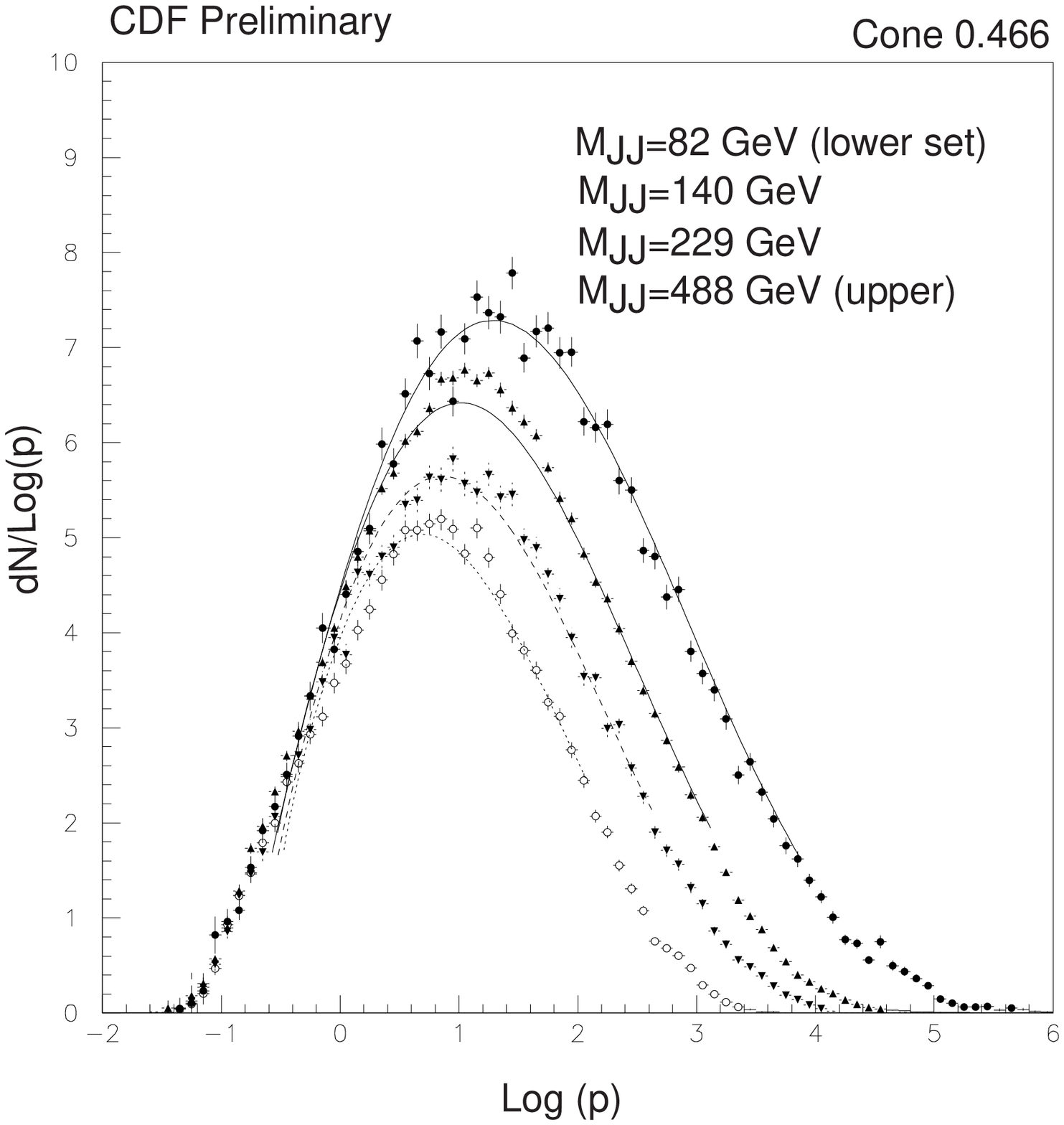} % postscript image file name
\caption{a) Differential cross--section in $\ln p$ for e$^+$e$^-$ data from 91 
to 200 \gev. b) The similar observable in $\rp\bar{\rp}$ collisions.
\label{dndp}}
\end{center}
\end{figure}

Since the scaled momentum may vail the effect with an absolute scale, 
it is also of interest to look at the differential cross-section for 
the unscaled momentum $E\rd n/\rd^3p$. In the low momentum range,
the calculation\cite{khoze} of this quantity exhibits the interesting 
feature of universality: just from looking at the low 
energetic part, one cannot tell if the annihilation took place at 50\gev, 
200 \gev\ or even higher energy.
This is understood as a consequence of the coherent emission of 
soft particles which cannot resolve the structure of the underlying event.
The same feature shows up also in $\rp\bar{\rp}$ collisions.
Fig.\ref{dndp} a) compares the DELPHI data from 91 to 200 \gev in 
$\rd n/\rd \ln p$ 
with the MLLA calculation, while Fig.\ref{dndp} b) shows the same quantity as 
measured by CDF\cite{cdf} for different invariant masses of the two-jet 
systems 
ranging from 80 to 500 \gev. Both data sets show the predicted universality 
in the soft part due to coherent emmission.
%%%%%%%%%%%%%%%%%%%%%%%%%%%%%%%%%%%%%%%%%%%%%%%%%%%%%%%%%%%%%%%%%%%
\section{Cone Multiplicity Perpendicular to the 3-Jet Plane}
\begin{figure}[ht]
\begin{center}
% \vspace{-0.5cm}
\epsfxsize=12pc % will enlarge or reduce the postscript figures based on t1he 
\epsfbox{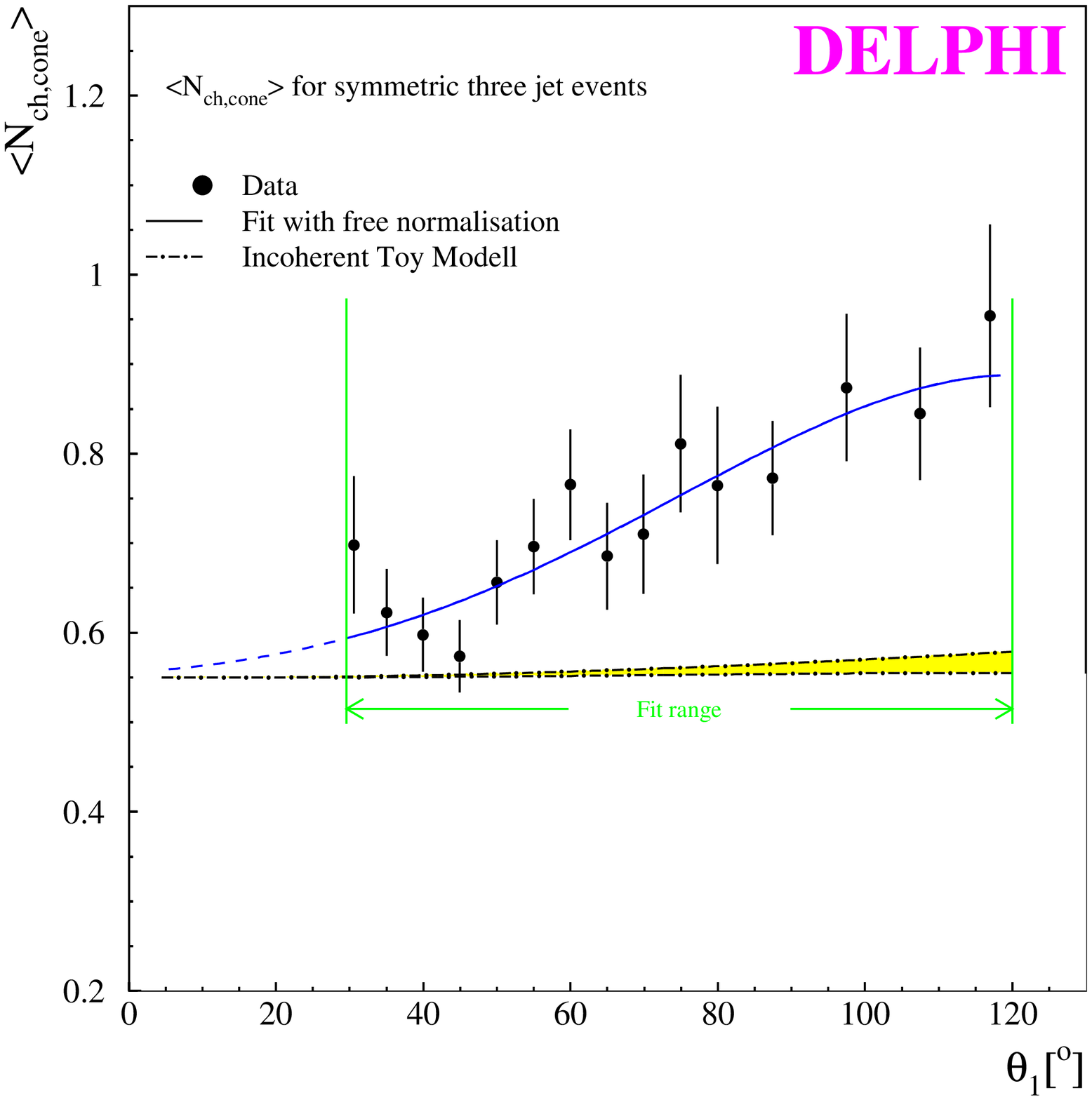} % postscript image file name
\epsfxsize=12pc % will enlarge or reduce the postscript figures based on t1he 
\epsfbox{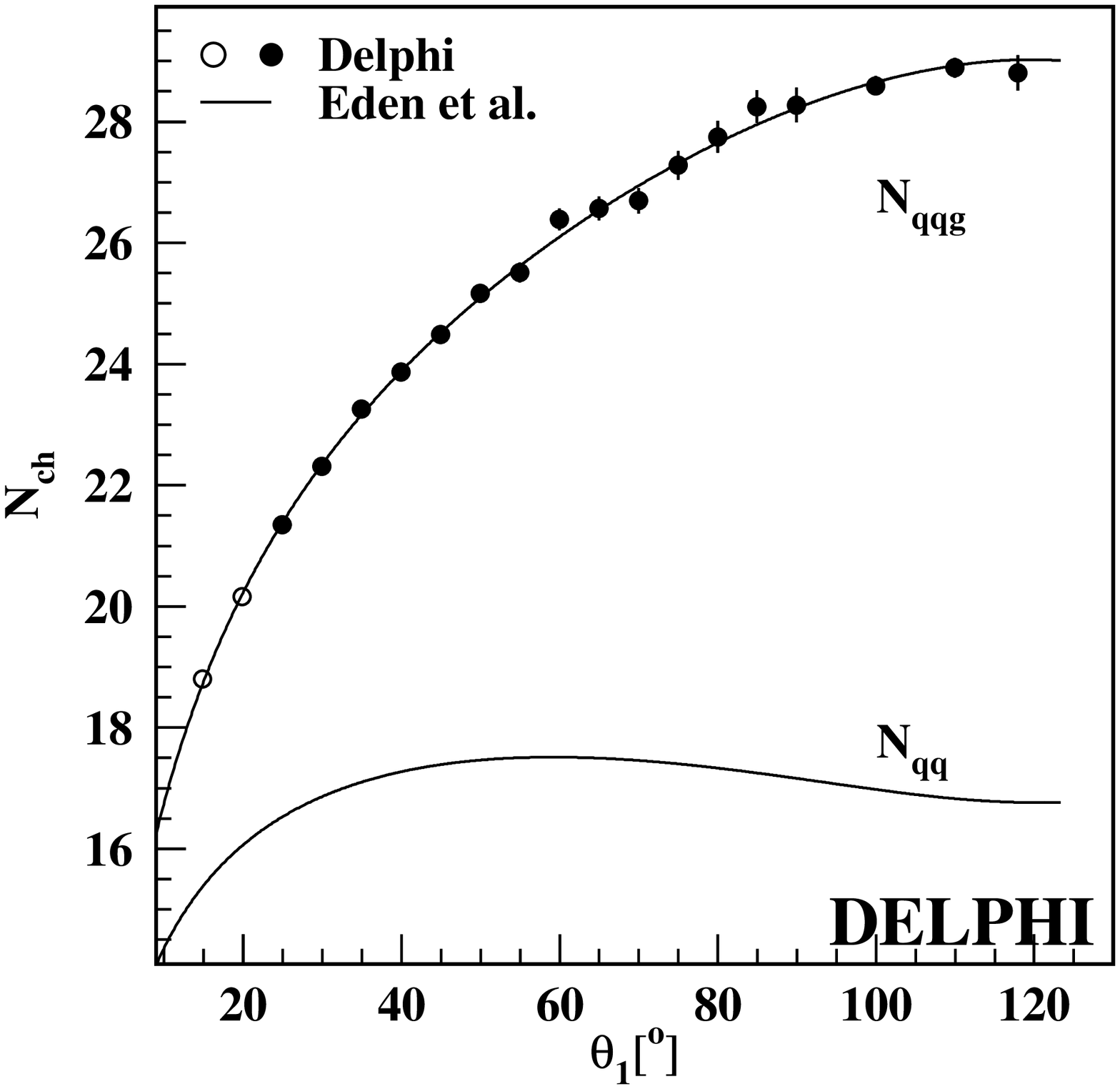} % postscript image file name
\caption{a) Cone multiplicity perpendicular to the event plane as a function 
of the  opening angle between the two less energetic jets. b)
 Multiplicity in symmetric three jet events as a
 function of the angle between the two less energetic jets. The full line is
 a fit of the Eden et al. prediction as discussed in the text. The phase space
 restricted $\rq\bar{\rq}$ contribution is displayed as the lower line.
\label{ochs}}
\end{center}
\end{figure}

Another test for coherent emission is provided by measuring the multiplicity 
perpendicular to the event plane in a three-jet event. This quantity has been
calculated\cite{khoze} as a function of the opening angles between the 
three jets. 
The physics behind this observable is the observation that, depending on the
opening angle between gluon and quark jet, the extra colour charge of the gluon
gets screened more or less. The DELPHI analysis \cite{delphi3} uses 
symmetric three-jet 
events, which makes it essentially unnecessary to identify the gluon jet.
Additionally, the formula simplifies to the expression:
$$
N_{\perp}^{\rq\bar{\rq}\rg}\sim 
\left( 2+\cos\frac{\theta_1}{2}-\cos \theta_1 - 
\frac{1}{N_C^2}(1+\cos\frac{\theta_1}{2}) \right )\ ,
$$
where $\theta_1$ is the angle between the two low-energetic jets. Fig. 
\ref{ochs} a) shows the data compared to this prediction. Indeed the 
measurment is in excellent agreement with the prediction.
%%%%%%%%%%%%%%%%%%%%%%%%%%%%%%%%%%%%%%%%%%%%%%%%%%%%%%%%%%%%%%%%%%%%
\section{Multiplicity in 3-Jet Events: $C_\rA/C_\rF$}
Of course there are more obvious things one can do with three jet 
events, like  accesing the colour factor ratio $C_\rA/C_\rF$, and thus the 
fundamental difference between gluons and quarks. The DELPHI analysis
for this measurment uses as starting point the following  
equation,\cite{eden} which relates the scale dependence of gluon and quark 
multiplicities:
$$
\frac{\rd N_{\rg\rg}(L^{\prime})}{\rd L^{\prime}}\sim\frac{N_\rC}{C_\rF}
\frac{\rd N_{\rq\bar{\rq}}}{\rd L}
$$
with $L=\ln{s/\Lambda^2}$ and $L^{\prime}=L+11/6-3/2$. Thus, by knowing the
multiplicity in $\rq\bar{\rq}$ events, the multiplicity in gluon--gluon 
events can 
be derived. In order to fix the boundary condition for this differential 
equation, CLEO data on $\Upsilon$ dacays can be used. This constant of 
integration takes  essentially non--perturbative effects into account.
Finally, the multiplicity in three jet events can be expressed as:
$$
N_{\rq\bar{\rq}\rg}=\frac{1}{2}N_{\rg\rg}(\kappa_{\Le})+N_{\rq\bar{\rq}}
(L_{\rq\bar{\rq}},\kappa_{\Lu})
$$
with $L_{\rq\bar{\rq}}=\ln{s_{\rq\bar{\rq}}/\Lambda^2}$, 
$\kappa_{\Lu}=\ln{s_{\rq\rg}
s_{\bar{\rq}\rg}/s\Lambda^2}$ and $\kappa_{\Lu}=\ln{s_{\rq\rg}
s_{\bar{\rq}\rg}/s_{\rq\bar{\rq}}\Lambda^2}$. 
The quark multiplicity entering in 
this expression is not exactly the directly measured one, but a ``phase-space 
restricted'' $\rq\bar{\rq}$ multiplicity. This seemingly incoherent sum of the
two contributions takes the coherence effect into account by a proper choice of
the evolution scales $\kappa$.

For the test of this prediction DELPHI\cite{delphi4} uses again symmetic 
three-jet events, thus that the whole event is characterized by one angle 
only. The multiplicity of 
the {\em whole} events as a function of this angle is now predicted as 
a function of four  quantities:   
The known $\rq\bar{\rq}$ multiplicities, the constant of integration from 
solving the differential equation (fixed by CLEO data as mentioned above),
 a multiplicity off--set $N_0$ for taking b quark effects into account
and the colour factor ratio $C_\rA/C_\rF$.

Note that only the {\em whole} event multiplicity is measured and any
unphysical subdivision into ``jet multiplicities'' is avoided.
  Fig. \ref{ochs} b) shows a fit of this prediction to the multiplicity in
symmetric three-jet events. For the two fitted parameters this yields 
(with statistical errors) $C_\rA/C_\rF=2.262\pm0.032$ and $N_0=0.760\pm0.047$.
\begin{figure}[ht]
\begin{center}
% \vspace{-0.5cm}
\epsfxsize=12pc % will enlarge or reduce the postscript figures based on t1he 
\epsfbox{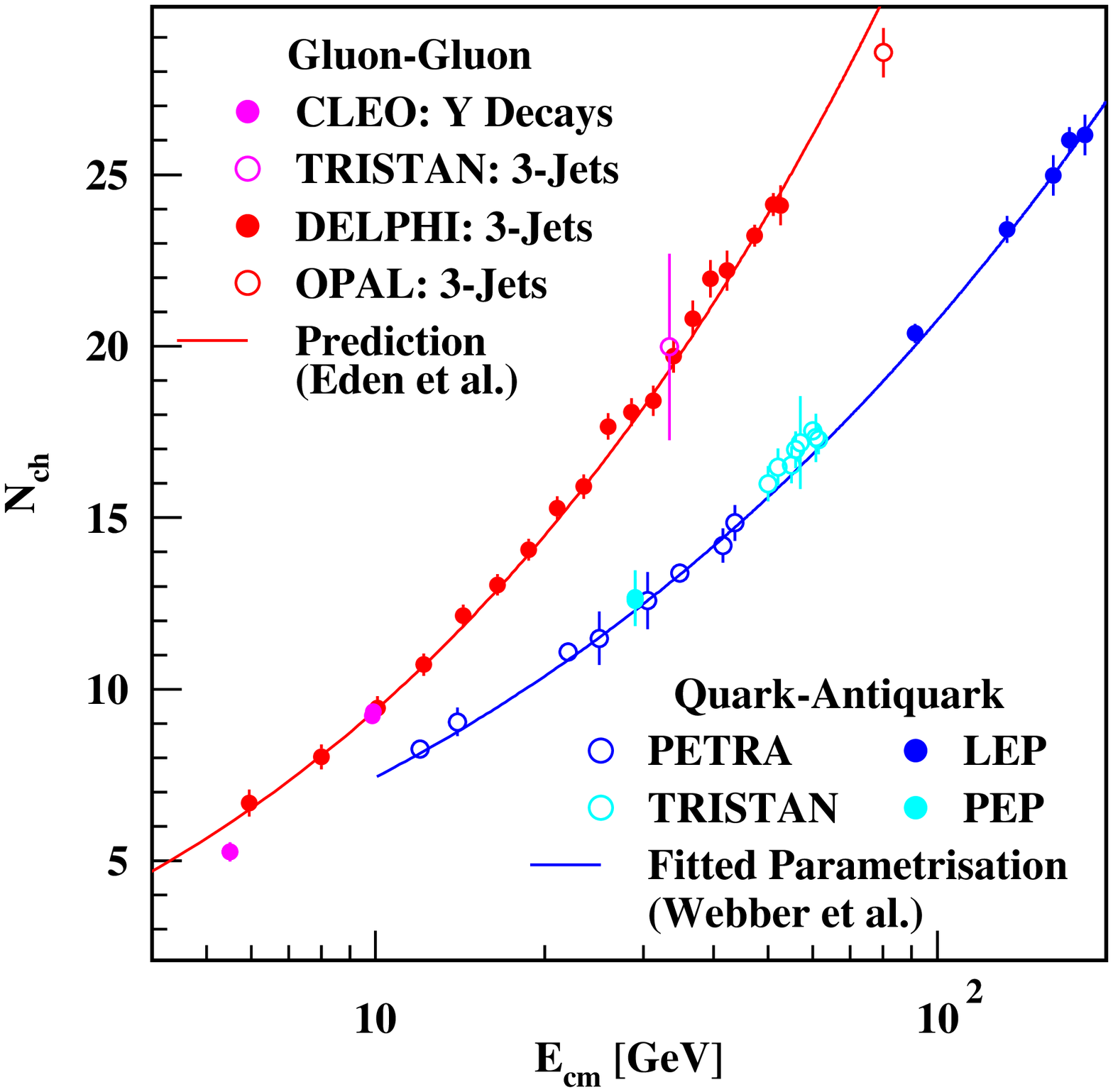} % postscript image file name
\epsfxsize=12pc
\epsfbox{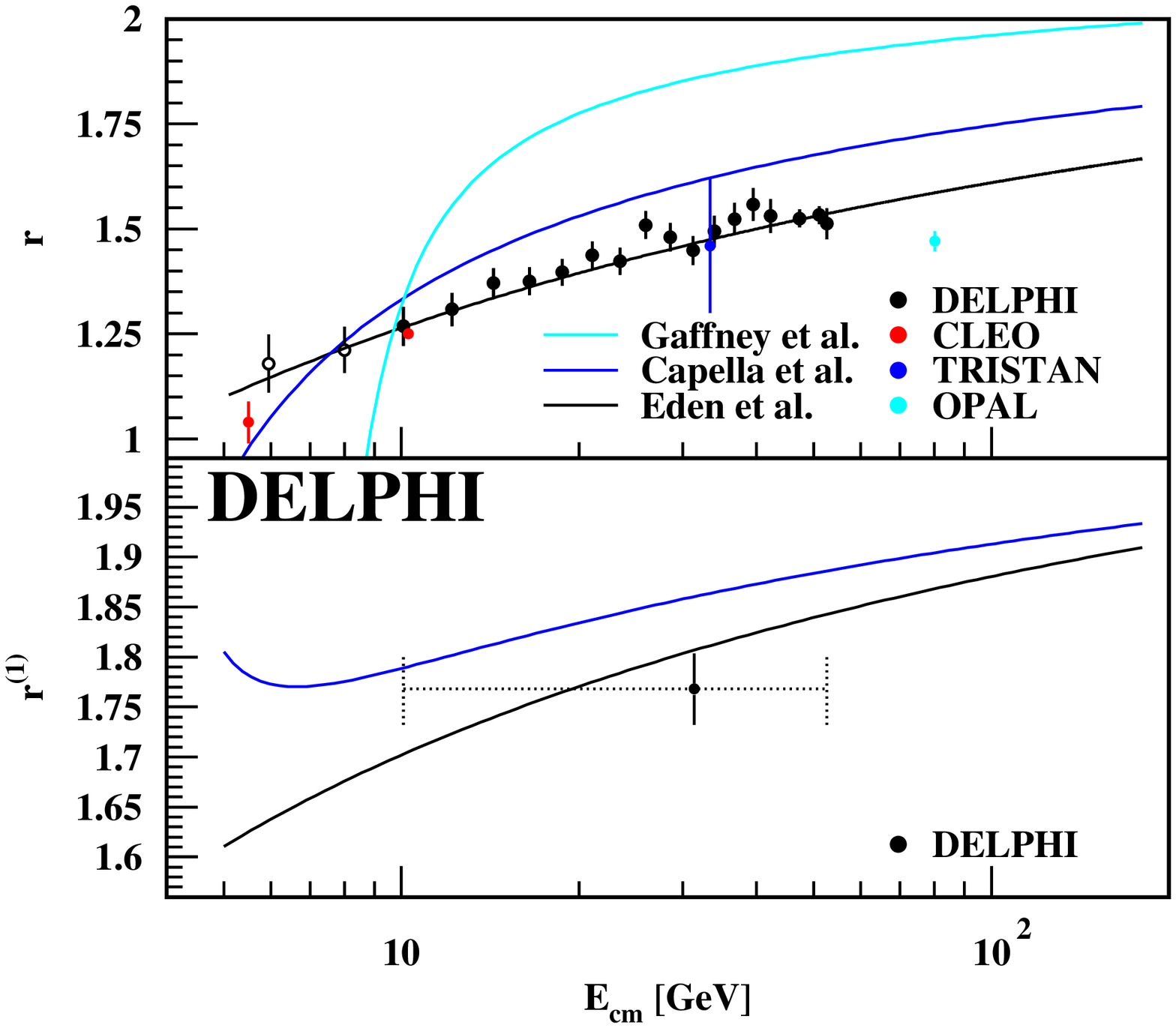} % postscript image file name
\caption{ a)  Multiplicity in $gg$ events as extracted from the measurement
 compared to quark multiplicities. b) gluon and quark multiplicity ratio 
(top)  and multiplicity-slope ratio (bottom).  \label{oliver}}
\end{center}
\end{figure}

By subtracting the quark contribution, one can get the multiplicity in 
gluon-gluon events, as shown in Fig. \ref{oliver} a), upper line. Here, the
angular dependence of the multiplicity is translated into the energy 
dependence according to the $p_\rt$ like $\kappa$ scale. The line
which describes the gluon data is this time not a fit, but the absolut 
prediction of the calculation.\cite{eden}
 The gluon multiplicities displayed in Fig.\ref{oliver}
are not only LEP1 results but also CLEO and TRISTAN measurements at 10 and
58 \gev, respectively.  
Fig.\ref{oliver} b) shows $r$, the ratio of gluon and quark multiplicities.
Its deviation from 1 was the first clear evidence for the bigger colour 
 charge of gluons, although it evidently does not provide a direct measure of 
$C_\rA/C_\rF$. It has also been calculated by other groups, 
as indicated in the 
plot,  but due to neglected non--perturbative effects these predictions
do not describe the data.
The lower plot in Fig.\ref{oliver} b) shows the slope--ratio 
$r^{(1)}$ of the multiplicities. Here, the deviation between the different 
calculations is less severe,  since the multiplicity slope is less affected 
by non--perturbative effects.
%\section{Summary}
%All QCD predictions for MPP show good agreement with the data, like the 
%charged  multiplicities, inclusive spectra and the multiplicity perpendicular
%to the event plane. The dead cone effect has been  confirmed experimentally. 
%The multiplicity of three jet events allows a precise determination of 
%$C_A/C_F$ if non--perturbative effect are taken into account. The lesson we 
%learn from that is that coherence effects are essential to describe 
%detailed properties of hadronic final staes.

\end{document}

%%%%%%%%%%%%%%%%%%%%%%%%%%%%%%%%%%%%%%%%%%%%%%%%%%%%%%%%%%%%%%%%%%%%%%%%%%%%%
%% End of  ws-p8-50x6-00.tex  
%%%%%%%%%%%%%%%%%%%%%%%%%%%%%%%%%%%%%%%%%%%%%%%%%%%%%%%%%%%%%%%%%%%%%%%%%%%%%